\documentstyle[preprint,aps,epsf]{revtex}
\begin{document}
\draft
\tightenlines
\preprint{IP/BBSR/95-23}
\title{
\bf{ PROTEIN FOLDING AND SPIN GLASS
} }
\author{ S. Suresh Rao\cite{eml1} and Somendra M.
Bhattacharjee\cite{eml} \\ }
\address{ Institute of Physics,
Bhubaneswar 751 005, India}
\maketitle
\begin{abstract}
We explicitly show the connection between the protein folding
problem and spin glass transition.  This is then used to
identify appropriate quantities that are required to describe
the transition.  A possible way of observing the spin glass
transition is proposed.
\end{abstract}


If a protein has to explore all the possible configurations to
reach its biologically active form, then the time required would
be $\sim 10^{10}$ years compared to the real situation, which is
of the order of few milliseconds to few seconds \cite{lev}.
This is the Levinthal paradox, whose resolution seems to hinge
on the similarities between this biologically important problem
and the concept of a rugged free energy landscape of a spin
glass in condensed matter physics \cite{dillchan,funnel}.
Attempts have also been made to study the dynamics directly but
so far these are necessarily restricted to small chains. Our
purpose in this paper is to establish the connection with the
spin glass quantitatively, and thereby identify the appropriate
quantities that one should look at in experiments.

The first idea that there is a connection with spin glass came
from the attempts to use a hierarchical tree structure of time
scales~\cite{frau}, and the random energy model~\cite{der} to
rationalize the observed dynamics of proteins~\cite{brywol} .
The connection was made apparent by the seminal work of Garel
and Orland \cite{go}, and, independently, of Gutin and
Shaknovich \cite{gushaka}.  From these evolved the idea of {\it
statistical proteins}. The key points in this approach are that
proteins are not simple homopolymers and functionally similar
proteins of different species need not have identical backbone
structure. The variation in an ensemble of such similar proteins
can be thought of as random (albeit correlated) sequence of
monomers along the backbone.  This randomness leads to random
interactions among the monomers.  Since the monomers do not
change positions once fixed, one has to consider averaging of
physical properties over the random realizations of monomer
configurations (quenched averaging). There will be quantities
whose average over the ensemble will be same as that of typical
samples, while there will be ones for which this is not true.
The former represent the class of quantities that are called
``self-averaging".  This class would represent the generic
properties of the proteins while the second class of
non-self-averaging quantities would be specific to samples
(species). The second class is expected to play a significant
role in mutants.

Granted the idea of statistical proteins, continuum path
integral formulations were used in Ref \cite{go,gushaka} to
calculate physical properties by using the replica theory.  In
approximate calculations, similarities with the infinite state
Potts glass were noted.  In particular, Ref \cite{go} shows the
importance of a finite bond length for a sensible theory.  It
was soon realized that the monomers in a real protein are not
just distributed at random, but there is a correlation, at least
to some extent, and the ensemble should be suitably restricted.

We take a model of finite bond lengths and exploit the
correlation to choose a certain combination of variables as the
independent random entities.  Unlike the previous approaches, we
use the bonds as the natural variables instead of the absolute
coordinates of the monomers.  We first establish that for
correlated distribution of monomers, the problem can be mapped
to a spin glass problem with long range interaction whose nature
is determined by the correlation.  We then identify the
parameters that would describe the spin glass state of the
protein, and this parameter is different from the measures of
size of homopolymers. For the particular type of correlations
considered, we obtain the exact scaling behaviour with the
length.  We then discuss how the parameter can be measured and
the existence of a spin glass phase can be verified.  The
relevance to dynamics is also discussed.

Let us start with the Kuhn model for a polymer consisting of
bonds of unit length freely joined at ends so that each can have
complete free rotation without any hindrance \cite{doied}.  See
Fig 1. We consider the problem in $d$ dimensions so that the
orientation of each bond is given by a $d$-dimensional vector
${\bf s}_p$, with $p$ going from $1$ to $N$, the total number of
bonds. Since the polymer configurations can be completely
specified by the $N$ direction vectors, we can as well consider
the equivalent problem of $d$ component spins arranged in one
dimension, representing the one dimensionality of the chain. It
is in this picture we formulate the problem.

Any two monomers $p$ and $q$ interact on contact with a coupling
proportional to $\epsilon_{pq}$, where $\epsilon_{pq}$ is a
quenched random variable. For a contact potential, this
interaction is $ \epsilon_{pq} \delta({\bf r}_{pq})$, where
${\bf r}_{pq}$ is the distance vector between monomers (sites)
$p$ and $q$, where $\delta(x)=0$ for $x\neq 0$ and $=1$ for
$x=0$.  In terms of the bond vectors, ${\bf r}_{pq} =
\sum_{i=p}^{q-1} {\bf s}_i$, the hamiltonian can be written as
\begin{equation}
  H=\sum_{p,q} \ \epsilon_{pq} \delta (\sum_{p}^{q-1}{\bf s}_i/(q-p)).
\end{equation}
This is the random version of the Domb-Joyce model for self
avoiding walk, and a positive $\epsilon_{pq}$ would represent a
repulsive (self avoiding term)\cite{domjoy}. We replace this
contact delta potential $\delta({\bf R})$ by a smoother
potential $1-R^2$.  Note that for the Kuhn model, $0<\mid {\bf
R}\mid<1$, so that the replacement is equivalent to changing the
discrete level 0 and 1 to a band between 0 and 1.  This leads to
much simplification afterwards. In a sense, two monomers
interact with a truncated quadratic potential that can be
repulsive or attractive. It is truncated just because the
distance between the two cannot exceed a limit - a feature of
the Kuhn model.  The proposed Hamiltonian is therefore
\begin{equation}
H=\sum_{ij} \ \frac{\epsilon_{ij}}{(j-i+1)^2}
\sum_{p,q \in [i,j)}  {\bf s}_p \cdot {\bf s}_q,\label{eq:ham2}
\end{equation}
ignoring a constant (disorder dependent) term that does not
contribute to the thermodynamics.  The Hamiltonian in this form
involves two sums.  Given a pair $i < j$, that defines a cluster
in the one dimensional chain, the inner sum involves a summation
over all the pairs in the cluster.  Let us rearrange the terms
and do the outer sum first.  For a pair $p,q$, sum over all the
clusters to which it belongs, to obtain terms of the type
$\sum_{{i\leq p}; {j\geq q}} \epsilon_{ij} (j-i+1)^{-2}$.  The
correlation of the monomers along the backbone (i.e. of
$\epsilon_{pq}$) is now invoked to write this sum over $i$ and
$j$ terms of independent random elements.  Specifically we
choose,
\begin{equation}
\sum_{m,n} \epsilon_{p-n,q+m} [(q-p) +(m-n)]^{-2} =
\frac{J_{pq}}{(q-p)^{\sigma}},\label{eq:sigma}
\end{equation}
with $J_{pq}$ as the independent random variable and the
exponent $\sigma$ as a measure of the correlation.  It is not
necessarily true that all correlations can be expressed in terms
of such a simple form, but this is the simplest situation.  More
complex situations can be handled by considering correlated,
and, if necessary, inhomogeneously distributed, $J_{pq}$.  This
does not invalidate the basic concepts introduced here.  The
ensemble we will be considering involves polymers that have a
particular type of correlations, as given by the distribution of
the couplings and the value of $\sigma$.

The Hamiltonian now takes a form familiar in the spin glass
context, namely
\begin{equation}
H= \sum_{p,q} \ \frac{J_{pq}}{(q-p)^{\sigma}} \ {\bf s}_p \cdot
{\bf s}_q, \label{eq:ham3}
\end{equation}
where each $J_{pq}$ is an independent normal variable with a
distribution $P(J_{pq}) = (2\pi J)^{-1/2}$ $\exp(-J_{pq}^2/2J)$.
This as a spin glass model is a generalization of the long range
Ising model considered in Ref.
\cite{kot} to vector spins\cite{craig}.

A spin glass transition is described in the $h,T$ plane where
$h$ is the magnetic field that orients the spins in a particular
direction.  The thermodynamic transition is heralded by a
diverging spin glass susceptibility, $\chi_{SG}$, while the
uniform, linear susceptibility, $\chi$, remains finite at the
transition\cite{fisch}.  One sees a cusp in $\chi$ at $T_c$.  In
terms of the correlation functions, the two susceptibilities can
be written as
\begin{equation}
\chi=N^{-1} \sum_{ij} {\overline {\langle {\bf s}_i\cdot {\bf
s}_j\rangle}} , \quad {\rm and} \quad
\chi_{SG}=N^{-1} \sum_{ij} {\overline {\langle {\bf s}_i\cdot {\bf
s}_j\rangle^2}} .
\end{equation}
We shall restrict ourselves to the high temperature disordered
phase, so that no special direction need be chosen. ( In
general, one should discuss the longitudinal and transverse
correlations \cite{craig}.) The important point to keep in mind
is the extensivity of the two susceptibilities, i.e., the total
susceptibilities (both linear and spin glass) are proportional
to the number of spins, so that the densities defined above are
independent of $N$.  In addition to the divergence of $\chi_{SG}
\sim \mid T- T_c\mid^{-\gamma}$, there is also a diverging
correlation length $\xi \sim \mid T- T_c\mid^{-\nu}$ which
describes the behavior of the correlation function
$g_{ij}={\overline {\langle {\bf s}_i\cdot {\bf
s}_j\rangle^2}}$.  The decay of the correlation at $T_c$ is
described by the exponent $\eta$, $g_{ij} \sim \mid j-i\mid
^{-1+\eta}$.  The response of the spin glass to an external
field can be written as $m = \chi h +\chi_{nl} h^3$, where ${\bf
m} = N^{-1} \sum_i {\overline {\langle {\bf s}_i\rangle}}$ is
the net magnetization in the field.  For symmetric
distributions, it is known that ({\it a}) the nonlinear
susceptibility $\chi_{nl}$ is related to the spin glass
susceptibility $\chi_{SG}$, a relation that is often used to
infer $\chi_{SG}$ from experiments \cite{chal}, and ({\it b})
only the diagonal correlations contribute to $\chi$.

We now translate these spin glass quantities to polymers.  The
spins in our problem correspond to the bonds of the polymer, so
that the total magnetization ${\bf M}=\sum_i {\bf s}_i$
corresponds to the end-to-end distance of the polymer.  This is
the quantity of interest in pure problems \cite{doied}.  Unless
the polymer is in a stretched state, the configurational average
of ${\bf M}$ is expected to be zero, and the size $R$ of the
polymer is given by the mean square end-to-end distance.  The
susceptibility is given by the variance of ${\bf M}$, and so,
with zero net magnetization, $\chi = \langle M^2\rangle/N$.  The
linear susceptibility of the spin system is therefore related to
the size of the polymer.  Since, as a density, $\chi$ is
independent of $N$, we find $R \sim N^{1/2}$, a result wellknown
from random walk. Remember that we are ignoring self avoidance -
that's why the random walk exponent.  In the spin glass case,
$\chi$ remains finite for all $T$, and, therefore, the size as
measured by $R$ in our model will always be proportional to
$N^{1/2}$, except that the temperature dependence in the strict
thermodynamic limit will show a singularity.  In contrast, close
to the transition temperature $\chi_{SG}$ shows a different
behavior.  A finite size scaling analysis \cite{bar} gives,
$\chi_{SG} \sim N^{\gamma/\nu}$, while away from $T_c$ it
remains $O(1)$.

We, therefore, propose that for the folding problem the
appropriate quantity to look at is
\begin{equation}
\Phi= \sum_{ij} {\overline {\langle {\bf s}_i\cdot {\bf
s}_j\rangle^2}}, \label{eq:phi}
\end{equation}
which goes like $\sim N^{1+\gamma/\nu}$ in the critical region,
but like $N$ for $T>>T_c$.  This is a different measure of size
than conventionally used in pure problems. Its importance can be
understood in terms of dynamics to be discussed below.  It is
possible to connect this $\Phi$ to the size of the polymer in
the following way:
\begin{equation}
{\overline{\langle R^2\rangle^2}} \sim
\sum {\overline{\langle {\bf s}_i \cdot {\bf s}_j\rangle
\langle{\bf s}_p \cdot {\bf s}_q\rangle }},
\end{equation}
and if we assume the dominance of the diagonal terms then,
\begin{equation}
\Phi \sim {\overline{\langle R^2\rangle^2}}.
\end{equation}
This is a justifiable assumption, since we do not require any
new exponent to describe the spin glass. Similar scaling is
expected for the radius of gyration also.  This gives an
experimentally accessible quantity that can be probed in
scattering experiments (see below).

Let us now go back to our Eq. \ref{eq:ham3}.  The spin glass
problem can be studied in the replica framework following the
method of Ref. \cite{kot}. Details are skipped.  The relevant
results we need here are the following: (1) There is a spin
glass transition for $1/2\leq\sigma< 1$. (2) For $\sigma<2/3$,
the behavior is meanfield like, and can as well be described by
an infinitely weak infinite range model \cite{craig}.  (3)
Fluctuations play a major role for $\sigma >2/3$ and the one
dimensional problem is expected to behave like a short ranged
spin glass.

As already pointed out after Eq. \ref{eq:phi}, the behaviour we
want to see comes from $\gamma/\nu$ which, by a scaling
relation, is equal to $2-\eta$ \cite{fisch}. Our aim is
therefore to calculate $\eta$.  Now, long range interactions do
not require any renormalization \cite{kot}. As a result, the
exponent $\eta$ is known exactly to be $\eta=3-2\sigma$. Hence,
the behavior of the fold parameter in the simple model is
determined as
\begin{eqnarray}
\Phi &\sim N^{2 \sigma} \quad {\rm for} \quad T\approx T_c
\nonumber \\
&\sim N \quad {\rm for } \quad T >>T_c,
\end{eqnarray}
for $2/3 <\sigma<1$. The restriction on $\sigma$ is needed
because finite size scaling is not valid for mean field
theories~\cite{brez}.  In other words, for $\sigma <2/3$, no
simple scaling form for $\Phi$ is expected near the transition.

A direct way of measuring the fold parameter $\Phi$ is to device
an experiment that stretches the polymer.  In the spin glass
language, the external field tries to orient all the spins along
its direction.  This ordered state corresponds to a stretched
rod-like configuration of the polymer.  The analog of the
magnetic field in spin glass is therefore a stretching force as
can be obtained by pulling the polymer at two ends (say by
putting tunable charges at the ends), or in extensional flows
that lead to a coil-stretch transition.  It is therefore
suggested that to elucidate the spin glass type behavior, it is
necessary to study the response of proteins in the glassy state
to a (may be oscillatory) stretching force, and look for the
nonlinear response.

Another way of measuring $\Phi$ would be to look at the
structure factor, especially in the leading correction (in
momenta) to the small angle scattering.  Let us for simplicity
assume that optically the protein behaves as a homopolymer,
i.e., in scattering, all the monomers behave identically, and
the thermal averaging can be approximated by a gaussian average.
The structure factor (see, e.g., Ref \cite{doied}) for a given
realization of the polymer is then given by $\exp( - k^2 R_g^2)$
for a wavevector ${\bf k}$, where $R_g^2= N^{-2} \sum_{m>n}
\langle({\bf r}_m - {\bf r}_n)^2\rangle$ is the square radius of
gyration.  A disorder averaged structure factor then gives
${\overline {R_g^2}} \sim \chi$ as the leading term in small
angle scattering (i.e., ${\bf k} \rightarrow 0$.) The next
correction depends on ${\overline {(R_g^2)^2}}$ which we argued
to have the same scaling behavior as $\Phi$.

So far we focussed on the equilibrium aspect of the problem.
The important time dependent activities (i.e. biological
functions) involve rearrangements (release of strains) through a
sequence of functionally important motions (FIMs). FIMs are the
movements of certain segments of the molecules involving or
surrounding the active site.  In our bond picture, the motion of
a block from $i$ to $j$ can be executed by an interchange of the
two spins $s_i$ and $s_j$.  For example, nearest neighbor $i,j$
interchange corresponds to the Verdier-Stockmeyer type
moves~\cite{doied} while the next nearest neighbor interchange
corresponds to a crankshaft motion~\cite{chandill}.  The FIMs
can then be identified as two spin interchanges (blocks
containing the active site), and one needs to classify them
according to time scales.  The relevant quantity to describe
such motions in the native state is to look at the time
correlation function
\begin{equation}
\Phi_1 (\tau)=  \lim_{N\rightarrow\infty} \langle
\{ {\bf s}_i(t) \cdot {\bf s}_j(t)\} \{ {\bf s}_i(t+\tau) {\bf
s}_j(t+\tau)\rangle\},\label{eq:phi1}
\end{equation}
where the average is now a time average.  For
$\tau\rightarrow\infty$, $\Phi_1(\infty)$ is the counterpart of
the Edwards-Anderson order parameter \cite{fisch} for spin
glasses.  The fold parameter $\Phi$ comes from Eq. \ref{eq:phi1}
if the limits are taken in the reverse order, i.e.,
$\lim_{N\rightarrow\infty}\lim_{\tau\rightarrow\infty} $.  It is
known that unlike $\Phi$, $\Phi_1(\infty)$, is not a
self-averaging quantity.

The importance of self-averaging quantities in the protein
context is that for such a quantity any typical sample behaves
like the average one.  In contrast, large sample to sample
fluctuations are expected in non-self-averaging quantities.  For
biological activity, mutants behave differently, mainly because
FIMs get modified.  It is, therefore, gratifying to find that
the measure $\Phi_1$ introduced above has the non-self-averaging
property that can distinguish a mutant or denatured protein from
the native one.

In summary, we have shown (in the spirit of lattice gas models
of liquid gas transition) that for a correlated heteropolymer,
the bonds variable are the suitable variables, and in terms of
these, the phase transition in the protein can be described by a
one dimensional vector spin glass model with long range
interaction.  We identified a fold parameter that should be the
measure for the folding problem.  Its exact scaling behavior
under certain circumstances has also been determined.  We find
exactly that for certain types of correlations, the geometric
exponents are determined completely by $\sigma$ of Eq.
\ref{eq:sigma}. The correlations along the backbone can
also destroy the scaling property, if $\sigma$ is large enough.
In other words, unlike the uncorrelated cases of Ref
\cite{go,gushaka}, our observations show that proteins need
not have a generic scaling behavior, and correlations do play a
major role in it.  We suggest that elastic moduli in oscillatory
stretching fields would help in the identification of the spin
glass type transition in proteins, {\it if there is one at all}.
Moreover, the proteins are inevitably of finite lengths, and
therefore what one can observe is not a true transition but the
finite size scaling behavior of the spin glass transition.  This
in turn opens up the new possibility of enriching our
understanding of spin glasses via controlled experiments done on
proteins with easily accessible $T_c$.

\begin{figure}
\begin{center}
\leavevmode
\hbox{%
\epsfxsize=5.0in
\epsffile{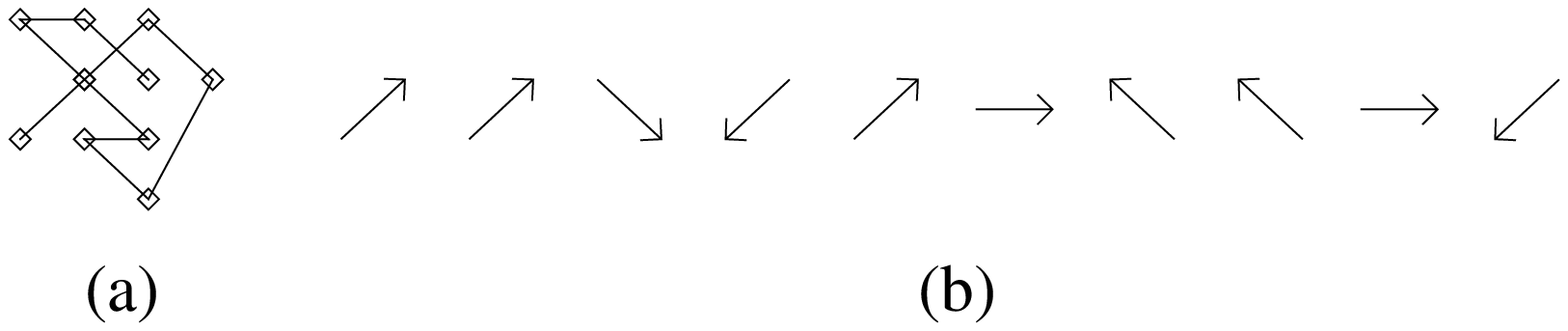}}
\end{center}
\caption{a) A segment of the Kuhn chain, and b) its one
dimensional spin representation.}
\end{figure}
\end{document}